\documentclass[a4paper,12pt]{article}
\usepackage{amsmath,amssymb,amsthm}

\newcommand{\bflambda}{{\boldsymbol \lambda}}
\newcommand{\bfzeta}{{\boldsymbol \zeta}}
\newcommand{\calB}{{\mathcal B}}
\newcommand{\calL}{{\mathcal L}}
\newcommand{\calM}{{\mathcal M}}

\newcommand{\calS}{{\mathcal S}}
\newcommand{\Complex}{{\mathbb C}}
\newcommand{\der}{\partial}

\newcommand{\fixed}{\text{fixed}}
\newcommand{\Res}{\operatorname{Res}\nolimits}
\newtheorem{theorem}{Theorem}[section]
\newtheorem{lemma}[theorem]{Lemma}
\newtheorem{proposition}[theorem]{Proposition}

\theoremstyle{definition}

\theoremstyle{remark}

\newtheorem{rem}[theorem]{Remark}
\newcommand\thmref[1]{Theorem~\ref{#1}}

\newcommand\secref[1]{\S\ref{#1}}

\begin{document}

\title{Radial L\"owner equation and \\Dispersionless cmKP Hierarchy}

\author{Kanehisa Takasaki
\\
Graduate School of Human and Environmental Sciences,\\ 
Kyoto University,\\
Kyoto 606-8501, Japan\\
E-mail: takasaki@math.h.kyoto-u.ac.jp
\\
\\
Takashi Takebe
\\
Department of Mathematics, Ochanomizu University\\
Otsuka 2-1-1, Bunkyo-ku, Tokyo, 112-8610, Japan\\ 
E-mail: takebe@math.ocha.ac.jp}

\date{}

\maketitle

\abstract{
It has been shown that the dispersionless KP hierarchy (or the Benney
hierarchy) is reduced to the chordal L\"owner equation. We show that the
radial L\"owner equation also gives reduction of a dispersionless type
integrable system. The resulting system acquires another degree of
freedom and becomes the dcmKP hierarchy, which is a ``half'' of the
dispersionless Toda hierarchy.


}

\section{Introduction}
\label{sec:intro}

Recently reductions and hodograph solutions of dispersionless and/or
Whitham type integrable systems are intensively studied
\cite{gib-tsa:99,yu-gib:00,m-a-m:02,g-m-a:03}.  In this article we
report another example --- reduction of the dispersionless coupled
modified KP (dcmKP) hierarchy to the (radial) L\"owner equation.

The L\"owner equation
\begin{equation}
    \frac{\der g}{\der \lambda}(\lambda,z)
    = g(\lambda,z) 
    \frac{\kappa(\lambda) + g(\lambda,z)}
         {\kappa(\lambda) - g(\lambda,z)}
      \frac{\der \phi(\lambda)}{\der \lambda}
\label{r-loewner:orig}
\end{equation}
was introduced by K.~L\"owner \cite{loe:23} in an attempt to solve the
Bieberbach conjecture. (See, e.g., \cite{dur:83}, Chapter 3.) It is an
evolution equation of the conformal mapping $g(\lambda,z)$ (as a
function of $z$) from a chain of subdomains of the complement of the unit
disk onto the complement of the unit disk, normalized as
\begin{equation*}
    g(\lambda,z) 
    = e^{-\phi(\lambda)} z 
    + b_0(\lambda) 
    + b_1(\lambda) z^{-1} + b_2(\lambda) z^{-2} + \cdots.
\end{equation*}
Namely, $g$ is normalized so that it maps a fixed interior point of the
domain ($z=\infty$) to a fixed interior point ($z=\infty$).

We can also define the same kind of equation with different
normalization which is called the ``{\em chordal} L\"owner equation'':
\begin{equation}
    \frac{\der g}{\der \lambda}(\lambda,z)
    = \frac{1}{g(\lambda,z) - U_i(\lambda)}
      \frac{\der a_1(\lambda)}{\der \lambda}.
\label{c-loewner:orig}
\end{equation}
Here $g$ is a conformal mapping from a subdomain of the upper half plain
to the upper half plain, normalized as
\begin{equation*}
    g(\lambda,z) 
    = z + a_1(\lambda) z^{-1} + a_2(\lambda) z^{-2} + \cdots.
\end{equation*}
Hence this maps a fixed boundary point ($z=\infty$) to a fixed boundary
point ($z=\infty$). See \cite{l-s-w:01} \S2.3 for details. The original
L\"owner equation is, therefore, often called the ``{\em radial}
L\"owner equation''.

The reduction of the dispersionless KP hierarchy
\cite{dkp,tak-tak:92,tak-tak:95} to the chordal L\"owner equation (and
its generalization) has been studied by Gibbons and Tsarev
\cite{gib-tsa:99}, Yu and Gibbons \cite{yu-gib:00}, Ma\~nas,
Mart\'{\i}nez Alonso and Medina \cite{m-a-m:02} and others. Our question
is: what about the radial L\"owner equation? Note that the above
mentioned works on the chordal case does not contain the radial L\"owner
equation because of the normalization at the infinity.

The answer is that there appears another degree of freedom and the
resulting system turns out to be the dcmKP hierarchy. The dcmKP
hierarchy introduced by Teo \cite{teo:03-1} is an extension of the
dispersionless mKP hierarchy \cite{tak:02} with an additional degree of
freedom, or in other words, a ``half'' of the dispersionless Toda
lattice hierarchy \cite{tak-tak:91,tak-tak:95}.

Ma\~nas, Mart\'{\i}nez Alonso and Medina \cite{m-a-m:02} generalized the
chordal L\"owner equation \eqref{c-loewner:orig} by changing the
rational function in the right hand side.  The radial L\"owner equation
is also generalized in the same way. In the theory of univalent
functions those generalized equations are called the L\"owner-Kufarev
equations. We show that a certain class of the L\"owner-Kufarev
equations is transformed to the (radial) L\"owner equation with respect
to the Riemann invariants. This class of L\"owner-Kufarev equations has
already been considered in \cite{pro-vas:04} and
\cite{car:03-05}. Prokhorov and Vasil'ev showed in \cite{pro-vas:04}
that such class is related to an integrable system, which is presumably
different from ours. Cardy introduced a stochastic version in
\cite{car:03-05}. 

In the following two sections we review the two ingredients, the
L\"owner equation and the dcmKP hierarchy. The main result is presented
in \S\ref{sec:loe->dcmkp}. The reduction of the L\"owner-Kufarev
equation is explained in \S\ref{sec:loe-kuf}. 

The results of this article was announced in \cite{tak-tak:05}.

\subsection*{Acknowledgements}
TT expresses his gratitude to A. Zabrodin for his interest and comments.
KT is partially supported by Grant-in-Aid for Scientific Research by
Japan Society for the Promotion of Science, No.\ 16340040.  TT is
partially supported by Grant-in-Aid for Scientific Research by Japan
Society for the Promotion of Science, No. 15540014.

\section{Radial L\"owner equation}
\label{sec:loewner}

In this section we review the (radial) L\"owner equation and introduce
related notions. Since we are interested in algebro-analytic nature of
the system, we omit reality/positivity conditions which are essential in
the context of the complex analysis.

The L\"owner equation is a system of differential equations for a
function 
\begin{equation}
    w= g(\bflambda,z) 
    = e^{-\phi(\bflambda)} z 
    + b_0(\bflambda) 
    + b_1(\bflambda) z^{-1} + b_2(\bflambda) z^{-2} + \cdots 
\label{g(lam,z)}
\end{equation}
where $\bflambda=(\lambda_1,\dots,\lambda_N)$ and $z$ are independent
variables. In the complex analysis the variable $z$ moves in a subdomain
of the complement of the unit disk and the variables $\lambda_i$
parametrize the subdomain. In our context $g(\bflambda,z)$ is considered
as a generating function of the unknown functions $\phi(\bflambda)$ and
$b_n(\bflambda)$. We assume that for each $i=1,\dots,N$ a {\em driving
function} $\kappa_i(\bflambda)$ is given. The {\em L\"owner equation} is
the following system:
\begin{equation}
    \frac{\der g}{\der \lambda_i}(\bflambda,z)
    = g(\bflambda,z) 
    \frac{\kappa_i(\bflambda) + g(\bflambda,z)}
         {\kappa_i(\bflambda) - g(\bflambda,z)}
      \frac{\der \phi(\bflambda)}{\der \lambda_i}, \qquad
    i=1,\dots,N.
\label{r-loewner:g}
\end{equation}
(The original L\"owner equation \eqref{r-loewner:orig} is the case $N=1$.)

Later the inverse function of $g(\bflambda,z)$ with respect to the
$z$-variable will be more important than $g$ itself. We denote it by
$f(\bflambda,w)$:
\begin{equation}
    z = f(\bflambda,w)
    = e^{\phi(\bflambda)} w 
    + c_0(\bflambda)
    + c_1(\bflambda) w^{-1} + c_2(\bflambda) w^{-2} + \cdots.
\label{f(lambda,w)}
\end{equation}
It satisfies $g(\bflambda,f(\bflambda,w)) = w$ and
$f(\bflambda,g(\bflambda,z)) = z$, from which we can determine the
coefficients $c_n(\bflambda)$ in terms of $\phi(\bflambda)$ and
$b_n(\bflambda)$. The L\"owner equation \eqref{r-loewner:g} is rewritten
as the equation for $f(\bflambda,w)$ as follows:
\begin{equation}
    \frac{\der f}{\der \lambda_i}(\bflambda,w)
    = w
    \frac{w + \kappa_i(\bflambda)}{w - \kappa_i(\bflambda)}
    \frac{\der \phi(\bflambda)}{\der \lambda_i}
    \frac{\der f}{\der w}(\bflambda,w).
\label{r-loewner:f}
\end{equation}
This equation leads to the compatibility condition of $\kappa_i$'s:
\begin{align}
    \frac{\der \kappa_j}{\der \lambda_i}
    &= - \kappa_j
    \frac{\kappa_j + \kappa_i}{\kappa_j-\kappa_i} 
    \frac{\der \phi}{\der \lambda_i},
\label{dk/dlambda}
\\
    \frac{\der^2 \phi}{\der\lambda_i \der\lambda_j}
    &=
    \frac{4 \kappa_i \kappa_j}{(\kappa_i - \kappa_j)^2}
    \frac{\der \phi}{\der \lambda_i}\, 
    \frac{\der \phi}{\der \lambda_j},
\label{d2phi/dlambda2}
\end{align}
for any $i,j$ ($i\neq j$).

The {\em Faber polynomials} are defined as follows \cite{teo:03-2}:
\begin{equation}
    \Phi_n(\bflambda,w) := (f(\bflambda,w)^n)_{\geq 0}.
\label{def:faber}
\end{equation}
Here $(\cdot)_{\geq 0}$ is the truncation of the Laurent series in $w$
to its polynomial part. (See also \cite{dur:83}, \S4.1.) The generating
function of $\Phi_n$'s is expressed in terms of $g(\bflambda,z)$:
\begin{equation}
    - \sum_{n=1}^\infty \frac{\Phi_n(w)}{n} z^{-n}
    = \log \frac{g(z) - w}{e^{-\phi} z}
    = \log (g(z) - w) + \phi - \log z.
\label{faber:gen}
\end{equation}
See \cite{dur:83} \S4.1 or \cite{teo:03-2} \S2.2. By differentiating
\eqref{faber:gen}, we have generating functions of derivatives of Faber
polynomials, which will be useful later:
\begin{equation}
\begin{aligned}
    - \sum_{n=1}^\infty \frac{1}{n} 
    \frac{\der \Phi_n}{\der w}(\bflambda, w) z^{-n}
    &=
    \frac{1}{w - g(\bflambda, z)},
\\
    - \sum_{n=1}^\infty \frac{1}{n} 
    \frac{\der^2 \Phi_n}{\der \lambda_i \der w}(\bflambda, w) z^{-n}
    &=
    \frac{g(\bflambda, z) (\kappa_i + g(\bflambda, z))}
         {(g(\bflambda, z) - w)^2 (\kappa_i - g(\bflambda, z))} 
    \frac{\der \phi}{\der \lambda_i},
\\
    - \sum_{n=1}^\infty \frac{1}{n} 
    \frac{\der^2 \Phi_n}{\der w^2}(\bflambda, w) z^{-n}
    &=
    - \frac{1}{(w - g(\bflambda, z))^2}.
\end{aligned}
\label{d(faber):gen} 
\end{equation}
The second equation is a consequence of the L\"owner equation
\eqref{r-loewner:g}. 

\section{dcmKP hierarchy}
\label{sec:dcmkp}

We give a formulation of the dcmKP hierarchy different from
Teo's \cite{teo:03-1}. The equivalence (up to a gauge factor) will be
explained in a forthcoming paper.

The independent variables of the system are
$(s,t)=(s,t_1,t_2,\dots)$. The unknown functions $\phi(s,t)$ and
$u_n(s,t)$ ($n=0,1,2,\dots$) are encapsulated in the series
\begin{equation}
    \calL(s,t;w) 
    = e^{\phi(s,t)} w 
    + u_0(s,t) + u_1(s,t) w^{-1} + u_2(s,t) w^{-2} + \cdots,
\label{def:L:dcmkp}
\end{equation}
where $w$ is a formal variable. The {\em dispersionless coupled modified
KP hierarchy} (dcmKP hierarchy) is the following system of differential
equations: 
\begin{equation}
    \frac{\der\calL}{\der t_n} = \{\calB_n, \calL\}, \qquad
    n=1,2,\dots.
\label{dcmkp}
\end{equation}
Here the Poisson bracket $\{,\}$ is defined by
\begin{equation}
    \{f(s,w),g(s,w)\} :=
    w \frac{\der f}{\der w} \frac{\der g}{\der s} -
    w \frac{\der f}{\der s} \frac{\der g}{\der w},
\label{def:poisson}
\end{equation}
and $\calB_n$ is the polynomial in $w$ defined by
\begin{equation}
    \calB_n
    := (\calL^n)_{>0} + \frac{1}{2} (\calL^n)_0,
\label{def:Bn:dcmkp}
\end{equation}
where $(\cdot)_{>0}$ is the positive power part in $w$ and $(\cdot)_0$
is the constant term with respect to $w$.

We use the following fact: Assume that there exists a function
$S(z,s,t)$ with the expansion
\begin{equation}
    S(z,s,t) 
    = \sum_{n=1}^\infty t_n z^n + s \log z - \frac{1}{2}\varphi
    - \sum_{n=1}^\infty \frac{v_n}{n} z^{-n},
\label{S-func:dcmkp}
\end{equation}
where $\varphi$ and $v_n$ are functions of $(s,t)$. A function $\calL =
\calL(s,t;w)$ of the form \eqref{def:L:dcmkp} is a solution of the dcmKP
hierarchy if and only if the following equation holds:
\begin{equation}
    dS(\calL,s,t)
    =
    \calM\, d\log\calL + \log w\, ds + \sum_{n=1}^\infty \calB_n dt_n,
\label{dS:dcmkp}
\end{equation}
where $\calM = \sum_{n=1}^\infty nt_n \calL^n + s + \sum_{n=1}^\infty
v_n \calL^{-n}$ is the Orlov-Schulman function and $\calB_n$ is defined
by \eqref{def:Bn:dcmkp}. In other words, $\calL$ is a solution if and
only if
\begin{equation}
    \left(\frac{\der S}{\der t_n}
    \right)_{z,s,t_m (m\neq n) \fixed, z=\calL}
    =
    \calB_n
\label{dS/dtn}
\end{equation}
and
\begin{equation}
    \left(\frac{\der S}{\der s}
    \right)_{z,t_n \fixed, z=\calL}
    =
    \log w
\label{dS/ds}
\end{equation}
hold. The proof is the same as the dKP or dToda case,
\cite{tak-tak:91,tak-tak:92,tak-tak:95}. Namely, we have only to note
that equation \eqref{dS:dcmkp} implies $d\omega = 0$ and $\omega
\wedge \omega = 0$ where $\omega:= d\log w \wedge ds + \sum_{n=1}^\infty
d\calB_n \wedge dt_n$.

\begin{rem}
 It follows from the constant term in \eqref{dS/dtn} and \eqref{dS/ds}
 that the function $\varphi(s,t)$ satisfies equations
 \eqref{varphi:cond} below.

 As is shown in \cite{teo:03-1}, a solution of the dispersionless Toda
 hierarchy is a solution of the dcmKP hierarchy if the half of the time
 variables are fixed. In this case the function $\varphi$ comes from the
 Toda field.
\end{rem}

\section{Reduction of dcmKP hierarchy to radial L\"owner equation}
\label{sec:loe->dcmkp}

In this section we show that a specialization of the variables
$\bflambda$ in $f(\bflambda,w)$ gives a solution of the dcmKP
hierarchy. 

Suppose $\bflambda(s,t) = (\lambda_1(s,t), \dots, \lambda_N(s,t))$
satisfies the equations
\begin{equation}
    \frac{\der \lambda_i}{\der t_n}
    = v^n_i(\bflambda(s,t))
    \frac{\der \lambda_i}{\der s}, \qquad i=1,\dots,N,
\label{lambda(s,t)}
\end{equation}
where $v^n_j(\bflambda)$ are defined by
\begin{equation}
    v^n_j(\bflambda)
    := \kappa_j(\bflambda)
    \frac{\der \Phi_n}{\der w}(\bflambda,\kappa_j(\bflambda))
    =
    \left.\frac{\der \Phi_n}{\der \log w} (\bflambda, w)
    \right|_{w=\kappa_j(\bflambda)}.
\label{def:vnj}
\end{equation}
Equations \eqref{lambda(s,t)} say that $\lambda_i(s,t)$ are Riemann
invariants with characteristic speed $v^n_i$.

\begin{lemma}
\label{lem:dvnj=Vij(vni-vnj)}
 The functions $v^n_j(\bflambda)$ satisfy the equations
\begin{equation}
    \frac{\der v^n_j}{\der \lambda_i}
    =
    V_{ij} (v^n_i-v^n_j),
\label{dv/dlambda}
\end{equation}
 where 
\begin{equation}
    V_{ij} :=
    \frac{2 \kappa_i \kappa_j}{(\kappa_i-\kappa_j)^2}
    \frac{\der \phi}{\der \lambda_i}.
\label{def:Vij}
\end{equation}
\end{lemma}

\begin{proof}
 By the chain rule and the generating function expressions
 \eqref{d(faber):gen}, we have
\begin{equation}
 \begin{split}
    - \sum_{n=1}^\infty \frac{1}{n}
    \frac{\der v^n_j}{\der \lambda_i} z^{-n}
    =&
    \frac{g}{(g-\kappa_j)^2}
    \left(
     - \frac{\der \kappa_j}{\der \lambda_i}
     + \frac{\kappa_j (g + \kappa_i)}{\kappa_i - g}
       \frac{\der \phi}{\der \lambda_i}
    \right)
\\
    =&
    \frac{g \kappa_j}{(g - \kappa_j)^2} \frac{\der \phi}{\der \lambda_i}
    \left(
      \frac{\kappa_j+\kappa_i}{\kappa_j-\kappa_i}
    + \frac{g+\kappa_i}{\kappa_i-g}
    \right)
\\
    =&
    - \frac{2 \kappa_i \kappa_j}{(\kappa_i - \kappa_j)^2}
    \frac{\der \phi}{\der \lambda_i}
    \left(\frac{\kappa_i}{g-\kappa_i} 
        - \frac{\kappa_j}{g-\kappa_j}\right)
 \end{split}
\label{dv/dlambda:gen}
\end{equation}
 Here we used \eqref{dk/dlambda}. The coefficient of $z^{-n}$ gives
 \eqref{dv/dlambda} because of \eqref{d(faber):gen}.
\end{proof}

The hydrodynamic type equations \eqref{lambda(s,t)} can be solved by the
generalized hodograph method of Tsarev \cite{tsa:91}:

\begin{proposition}
\label{prop:hodograph}
(i)
 For any triple of distinct indices $i,j,k$, we have
\begin{equation}
    \frac{\der V_{jk}}{\der \lambda_i}
    =
    \frac{\der V_{ik}}{\der \lambda_j}.
\label{compatible:Fi:cond}
\end{equation}
 Hence the system for functions $F_i(\bflambda)$
\begin{equation}
    \frac{\der F_j}{\der \lambda_i}
    =
    V_{ij} (F_i - F_j)
\label{Fi:cond}
\end{equation}
 is compatible according to \cite{tsa:91} \S3.

(ii)
 Let $\{F_i\}_{i=1,\dots,N}$ be a solution of \eqref{Fi:cond}. Then the
 {\em hodograph relation}
\begin{equation}
    F_i(\bflambda(s,t)) 
    = s + \sum_{n=1}^\infty v^n_i(\bflambda(s,t)) \, t_n
\label{hodograph}
\end{equation}
 defines a solution $\bflambda(s,t)$ of \eqref{lambda(s,t)} as an
 implicit function.
\end{proposition}

\begin{proof}
 (i) is a consequence of direct computation. In fact the both hand sides
 of \eqref{Fi:cond} are equal to
\begin{equation*}
    \frac{4 \kappa_i \kappa_j \kappa_k (\kappa_i+\kappa_j)}
         {(\kappa_i-\kappa_k)(\kappa_j-\kappa_k)(\kappa_i-\kappa_j)^2} 
    \frac{\der \phi}{\der \lambda_i}
    \frac{\der \phi}{\der \lambda_j},
\end{equation*}
 due to \eqref{dk/dlambda} and \eqref{d2phi/dlambda2}.

 (ii)
 The idea of the proof is the same as that of Theorem 10 of
 \cite{tsa:91}. By differentiating \eqref{hodograph} by $s$ and $t_n$,
 we obtain
\begin{align}
    \sum_{j=1}^N
    \left(
    \frac{\der F_i}{\der \lambda_j}(\bflambda(s,t)) 
    -
    \sum_{m=1}^\infty 
    \frac{\der v^m_i}{\der \lambda_j}(\bflambda(s,t))\, t_m
    \right)
    \frac{\der \lambda_j}{\der s}
    &=
    1,
\label{d(hodograph)/ds}
\\
    \sum_{j=1}^N
    \left(
    \frac{\der F_i}{\der \lambda_j}(\bflambda(s,t)) 
    -
    \sum_{m=1}^\infty 
    \frac{\der v^m_i}{\der \lambda_j}(\bflambda(s,t))\, t_m
    \right)
    \frac{\der \lambda_j}{\der t_n}
    &=
    v^n_i(\bflambda(s,t)).
\label{d(hodograph)/dtn}
\end{align}
 The first factor in the left hand side in \eqref{d(hodograph)/ds} and
 \eqref{d(hodograph)/dtn} with $j\neq i$ is
\begin{equation*}
 \begin{split}
    &\frac{\der F_i}{\der \lambda_j}(\bflambda(s,t)) 
    -
    \sum_{m=1}^\infty 
    \frac{\der v^m_i}{\der \lambda_j}(\bflambda(s,t))\, t_m
\\
    =&
    \frac{\der F_i}{\der \lambda_j}(\bflambda(s,t)) 
    -
    V_{ji}
     \sum_{m=1}^\infty \bigl(
     v^m_j(\bflambda(s,t))\, t_m - v^m_i(\bflambda(s,t)) t_m
     \bigr)
\\
    =&
    \frac{\der F_i}{\der \lambda_j}(\bflambda(s,t)) 
    - 
    V_{ji} \bigl(F_j(\bflambda(s,t)) - F_i(\bflambda(s,t)) \bigr)
    = 0,
 \end{split}
\end{equation*}
 because of \eqref{dv/dlambda}, \eqref{hodograph} and
 \eqref{Fi:cond}. Hence equations \eqref{d(hodograph)/ds} and
 \eqref{d(hodograph)/dtn} become
\begin{align*}
    \left(
    \frac{\der F_i}{\der \lambda_i}(\bflambda(s,t)) 
    -
    \sum_{m=1}^\infty 
    \frac{\der v^m_i}{\der \lambda_i}(\bflambda(s,t))\, t_m
    \right)
    \frac{\der \lambda_i}{\der s}
    &=
    1,
\\
    \left(
    \frac{\der F_i}{\der \lambda_i}(\bflambda(s,t)) 
    -
    \sum_{m=1}^\infty 
    \frac{\der v^m_i}{\der \lambda_i}(\bflambda(s,t))\, t_m
    \right)
    \frac{\der \lambda_i}{\der t_n}
    &=
    v^n_i(\bflambda(s,t)),
\end{align*}
 the ratio of which gives \eqref{lambda(s,t)}.
\end{proof}

Our main result is the following.

\begin{theorem}
\label{thm:loewner->dcmkp}
 Let $f(\bflambda,w)$ be a solution of the radial L\"owner equation
 \eqref{r-loewner:f} of the form \eqref{f(lambda,w)} and
 $\bflambda(s,t)$ be a solution of \eqref{lambda(s,t)}. Then the
 function $\calL=\calL(s,t;w)$ defined by 
\begin{multline}
    \calL(s,t;w) := f(\bflambda(s,t), w)
\\    = e^{\phi(\bflambda(s,t))} w 
    + c_0(\bflambda(s,t))
    + c_1(\bflambda(s,t)) w^{-1} + c_2(\bflambda(s,t)) w^{-2} + \cdots
\label{def:L}
\end{multline}
is a solution of the dcmKP hierarchy \eqref{dcmkp}.
\end{theorem}

The rest of this section is devoted to the proof of this theorem. We
construct the $S$ function \eqref{S-func:dcmkp}, following Ma\~nas,
Mart{\'\i}nez Alonso and Medina \cite{m-a-m:02}, but the zero-mode
$-\varphi/2$ in \eqref{S-func:dcmkp} should be added separately.

\begin{lemma}
\label{lem:zero-mode}
 There exists a function $\varphi(s,t)$ which satisfies
\begin{equation}
    \frac{\der \varphi}{\der t_n} = \Phi_n(\bflambda(s,t), 0), \qquad
    \frac{\der \varphi}{\der s} = 2 \phi(s,t),
\label{varphi:cond}
\end{equation}
 for all $n=1,2,\dots$.
\end{lemma}

\begin{proof}
 We have only to check the compatibility of \eqref{varphi:cond}, i.e.,
\begin{equation}
    \frac{\der \Phi_n}{\der t_m} (\bflambda(s,t), 0)
    =
    \frac{\der \Phi_m}{\der t_n} (\bflambda(s,t), 0),\qquad
    \frac{\der \Phi_n}{\der s} (\bflambda(s,t), 0)
    =
    2 \frac{\der \phi}{\der t_n}.
\label{varphi:compatibility}
\end{equation}
 Thanks to the generating function expression \eqref{faber:gen}, the
 left hand side of the second equation of \eqref{varphi:compatibility}
 is 
\begin{equation}
 \begin{split}
    &\frac{\der\Phi_n}{\der s}(\bflambda(s,t),0)
\\
    =& -n 
    \Res z^{n-1} \frac{\der}{\der s} 
    \left(
     \log g(\bflambda(s,t), z)
     + \phi(\bflambda(s,t)) - \log z
    \right)\, dz
\\
    =&
    - \Res \frac{\der}{\der s} \log g(\bflambda(s,t), z)
    \, d (z^n)
\\
    =&
    \Res z^n \,
    d \left(\frac{\der}{\der s} \log g(\bflambda(s,t), z)\right),
 \end{split}
\label{dPhi(0)/ds:temp}
\end{equation}
 by integration by parts. Since the L\"owner equation implies
\begin{equation}
 \begin{split}
    \frac{\der}{\der s} \log g(\bflambda(s,t), z)
    =&
    \sum_{i=1}^N \frac{\der \lambda_i}{\der s}
    \frac{1}{g(\bflambda(s,t),z)} \frac{\der g}{\der \lambda_i}
\\
    =&
    \sum_{i=1}^N \frac{\der \lambda_i}{\der s}
    \frac{\kappa_i + g}{\kappa_i - g} \frac{\der \phi}{\der \lambda_i},
 \end{split}
\label{dlogg/ds}
\end{equation}
 we can rewrite \eqref{dPhi(0)/ds:temp} by the coordinate transformation
 $w=g(\bflambda,z)$ (i.e., $z=f(\bflambda,w)$) as follows:
\begin{equation}
 \begin{split}
    &\frac{\der\Phi_n}{\der s}(\bflambda(s,t),0)
\\
    =&
    \Res f(\bflambda(s,t),w)^n \,
    \frac{\der}{\der w}
    \left(
     \left.\frac{\der}{\der s} \log g(\bflambda(s,t), z)
     \right|_{z=f(\bflambda(s,t),w)}
    \right)\, dw
\\
    =&
    \sum_{i=1}^N \frac{\der \lambda_i}{\der s}
    \Res \left( f(\bflambda(s,t),w)^n 
    \frac{\der}{\der w} \frac{\kappa_i + w}{\kappa_i - w}
    \frac{\der \phi}{\der \lambda_i} \right)\, dw
\\
    =&
    - \sum_{i=1}^N \frac{\der \lambda_i}{\der s}
    \frac{\der \phi}{\der \lambda_i} 
    \Res \left(
     f(\bflambda(s,t),w)^n \frac{- 2\kappa_i}{(w-\kappa_i)^2}
    \right)\, dw.
 \end{split}
\label{dPhi(0)/ds:temp2}
\end{equation}
 The argument which proves (2.5) of \cite{teo:03-2} shows that the
 residue in the last line is $-2\kappa_i\dfrac{\der \Phi_n}{\der
 w}(\bflambda(s,t),\kappa_i) = -2 v^n_i$. Hence it follows from
 \eqref{dPhi(0)/ds:temp2} that 
\begin{equation*}
    \frac{\der\Phi_n}{\der s}(\bflambda(s,t),0)
    =
    2 \sum_{i=1}^N v^n_i \frac{\der \lambda_i}{\der s}
    \frac{\der \phi}{\der \lambda_i} 
    =
    2 \sum_{i=1}^N \frac{\der \lambda_i}{\der t_n}
    \frac{\der \phi}{\der \lambda_i} 
    =
    2 \frac{\der \phi}{\der t_n},
\end{equation*}
 which proves the second equation of \eqref{varphi:compatibility}. The
 first equation of \eqref{varphi:compatibility} is proved in the same
 manner.
\end{proof}

Using this $\varphi(s,t)$, we define the $S$ function as
\begin{gather}
    S(z,s,t) := \calS(g(\bflambda(s,t), z), \bflambda(s,t), s, t)
              - \frac{1}{2} \varphi(s,t),
\label{def:S}
\\
    \calS(w, \bflambda, s, t) := 
    \calS_+(w, \bflambda,s, t) + \calS_-(w, \bflambda),
\label{def:calS}
\end{gather}
where
\begin{gather}
 \begin{split}
    \calS_+(w, \bflambda,s,t) 
    := \sum_{n=1}^\infty t_n \Phi_n(\bflambda, w) 
                       + s \log e^{\phi(\bflambda)} w
\\
     = \sum_{n=1}^\infty t_n \Phi_n(\bflambda, w) 
                       + s (\phi(\bflambda) + \log w),
 \end{split}
\label{def:calS+}
\end{gather}
and $\calS_-$ is a power series of $w^{-1}$ without a constant term
(i.e., $\calS_-(w, \bflambda) = O(w^{-1})$) which satisfies the
differential equation:
\begin{equation}
    \frac{\der\calS_-}{\der \lambda_i} -
    w \frac{w+\kappa_i}{w-\kappa_i} \frac{\der\phi}{\der\lambda_i}
    \frac{\der\calS_-}{\der w}
    =
    \frac{2\kappa_i F_i}{w - \kappa_i}
    \frac{\der\phi}{\der\lambda_i}.
\label{calS-:cond}
\end{equation}
The compatibility of \eqref{calS-:cond} is ensured by
\eqref{Fi:cond}. (cf.\ \cite{g-m-a:03} Proposition 2.)

Let us check that the $S$ function satisfies \eqref{dS/dtn} and
\eqref{dS/ds}. The argument is similar to the proof of Proposition 2 of
\cite{g-m-a:03}. First we prove \eqref{dS/dtn}. By the definition
\eqref{def:S}, we have
\begin{equation}
 \begin{split}
    \frac{\der S}{\der t_n}
    =&
    \left.\frac{\der\calS}{\der t_n}(w,\bflambda,s,t)
    \right|_{w=g(\bflambda(s,t),z), \bflambda=\bflambda(s,t)}
\\
    +&
    \sum_{i=1}^N 
    \left.\frac{\der}{\der \lambda_i}
    \calS(g(\bflambda,z), \bflambda, s, t)
    \right|_{\bflambda=\bflambda(s,t)}
    \frac{\der\lambda_i}{\der t_n}
    - \frac{1}{2} \frac{\der\varphi}{\der t_n}.
 \end{split}
\label{dS/dtn:temp}
\end{equation}
The first term of the right hand side can be written as
\begin{equation}
 \begin{split}
    \left.\frac{\der\calS}{\der t_n}(w,\bflambda,s,t)
    \right|_{w=g(\bflambda(s,t),z), \bflambda=\bflambda(s,t)}
    =& \Phi_n(\bflambda(s,t), g(\bflambda(s,t),z))
\\
    =& \Phi_n(\bflambda(s,t), w)
 \end{split}
\label{dS/dtn:1}
\end{equation}
by the definitions \eqref{def:calS} and \eqref{def:calS+}. We shall show
that 
\begin{equation}
     \left.\frac{\der}{\der \lambda_i}
      \calS(g(\bflambda,z), \bflambda, s, t)
     \right|_{\bflambda=\bflambda(s,t)}
     =0,
\label{dS/dlambdai=0}
\end{equation}
which proves \eqref{dS/dtn} thanks to \eqref{varphi:cond}.

We divide \eqref{dS/dlambdai=0} into two parts, namely,
\begin{equation}
    \left(
     \left.\frac{\der}{\der \lambda_i}
      \calS(g(\bflambda,z), \bflambda, s, t)
     \right|_{\bflambda=\bflambda(s,t)}
    \right)_{\geq 0}
    = 0,
\label{dS/dlambdai<0}
\end{equation}
and
\begin{equation}
    \left(
     \left.\frac{\der}{\der \lambda_i}
      \calS(g(\bflambda,z), \bflambda, s, t)
     \right|_{\bflambda=\bflambda(s,t)}
    \right)_{<0} = 0.
\label{dS/dlambdai>0}
\end{equation}
We first prove \eqref{dS/dlambdai<0}. Since $\calS_-$ consists of
negative powers of $w$, it suffices to show
\begin{equation}
    \left(
     \left.\frac{\der}{\der \lambda_i}
      \calS_+(g(\bflambda,z), \bflambda, s, t)
     \right|_{\bflambda=\bflambda(s,t)}
    \right)_{\geq 0}
    = 0.
\label{dS+/dlambdai<0}
\end{equation}
Recall that we have
\begin{equation*}
    \Phi_n(\bflambda,g(\bflambda,z)) = z^n + O(z^{-1}), \qquad
    \log e^\phi g(\bflambda,z) = \log z + O(z^{-1}),
\end{equation*}
by the definition of the Faber polynomials \eqref{def:faber} and the
normalization \eqref{g(lam,z)} of $g(\bflambda,z)$. Therefore
\begin{equation*}
 \begin{split}
    &\frac{\der}{\der \lambda_i} \calS_+(g(\bflambda,z),\bflambda,s,t)
\\
    =&
    \sum_{n=1}^\infty 
    t_n \frac{\der \Phi_n}{\der \lambda_i}(\bflambda,g(\bflambda,z))
    + s \frac{\der}{\der \lambda_i} \log e^{\phi}g(\bflambda,z)
    = O(z^{-1}),
 \end{split}
\end{equation*}
and hence the expression in the parentheses in \eqref{dS+/dlambdai<0}
does not contain non-negative powers of $w$, when $z$ is substituted by
$f(\bflambda,w)$.

Next we prove \eqref{dS/dlambdai>0}.  From the definition of $\calS_+$
\eqref{def:calS+} we can rewrite the left hand side of
\eqref{dS/dlambdai>0} as
\begin{equation}
 \begin{split}
    &\left(
     \left.\frac{\der}{\der \lambda_i}
      \calS(g(\bflambda,z), \bflambda, s, t)
     \right|_{\bflambda=\bflambda(s,t)}
    \right)_{<0}
\\  
    =&
    \left(
     \left.
      \frac{\der\calS}{\der w}(g(\bflambda,z), \bflambda, s, t)
      \frac{\der g}{\der \lambda_i}(\bflambda,z)      
     \right|_{\bflambda=\bflambda(s,t)}
    \right)_{<0}
\\
      &+
     \left.
      \frac{\der\calS_-}{\der \lambda_i}(g(\bflambda,z), \bflambda, s, t)
     \right|_{\bflambda=\bflambda(s,t)}
 \end{split}
\label{dS/dlambdai>0:temp1}
\end{equation}
Using the L\"owner equation, we have
\begin{equation}
 \begin{split}
    & \frac{\der\calS}{\der w}(g(\bflambda,z), \bflambda, s, t)
      \frac{\der g}{\der \lambda_i}(\bflambda,z)      
\\
    =& \frac{\der\calS}{\der w}(g(\bflambda,z), \bflambda, s, t) \,
      g(\bflambda,z) 
      \frac{\kappa_i(\bflambda) + g(\bflambda,z)}
         {\kappa_i(\bflambda) - g(\bflambda,z)}
    \frac{\der \phi}{\der \lambda_i}
\\
    =&
    \left(
      w 
      \frac{\der\calS_+}{\der w}(w, \bflambda, s, t) 
      - F_i(\bflambda)
    \right)
    \left.
    \frac{\kappa_i(\bflambda) + w}
         {\kappa_i(\bflambda) - w}
    \frac{\der \phi}{\der \lambda_i}
    \right|_{w=g(\bflambda,z)}.
\\
    &+ \left(
      w 
      \frac{\der\calS_-}{\der w}(w, \bflambda, s, t) 
      + F_i(\bflambda)
    \right)
    \left.
    \frac{\kappa_i(\bflambda) + w}
         {\kappa_i(\bflambda) - w}
    \frac{\der \phi}{\der \lambda_i}
    \right|_{w=g(\bflambda,z)}.
 \end{split}
\label{dS/dlambdai>0:temp3}
\end{equation}
Note that
\begin{multline}
    w \frac{\der\calS_+}{\der w}(w, \bflambda(s,t), s, t) 
    - F_i(\bflambda(s,t))
\\
    =
    s + \sum_{n=1}^\infty 
    t_n w \frac{\der\Phi_n}{\der w}(w;\bflambda(s,t),s,t) 
    - F_i(\bflambda(s,t)),
\label{wdS+/dw-F}
\end{multline}
which vanishes at $w=\kappa_i(\bflambda(s,t))$ because of the hodograph
relation \eqref{hodograph}. Hence
\begin{equation*}
    \left(
      w 
      \frac{\der\calS_+}{\der w}(w, \bflambda(s,t), s, t) 
      - F_i(\bflambda(s,t))
    \right)
    \frac{\kappa_i(\bflambda(s,t)) + w}
         {\kappa_i(\bflambda(s,t)) - w}
\end{equation*}
is an entire function of $w$ and, in particular, does not contain
negative powers of $w$ in its expansion at $w=\infty$. On the other
hand, we have
\begin{equation}
 \begin{split}
    &\left(
     \left(
       w
       \frac{\der\calS_-}{\der w}(w, \bflambda, s, t) 
       + F_i(\bflambda)
     \right)
     \frac{\kappa_i(\bflambda) + w}
          {\kappa_i(\bflambda) - w}
    \right)_{<0}
\\
    =&
    w \frac{\der\calS_-}{\der w}(w, \bflambda, s, t) 
    \frac{\kappa_i(\bflambda) + w}
         {\kappa_i(\bflambda) - w}
    + \frac{2\kappa_i(\bflambda)F_i(\bflambda)}
           {\kappa_i - w} = 0,
 \end{split}
\label{(wdS-/dw - Fi)<0}
\end{equation}
which follows from $\calS_- = O(w^{-1})$ and
\eqref{calS-:cond}. Therefore \eqref{dS/dlambdai>0} holds, which,
together with \eqref{dS/dlambdai<0}, leads to \eqref{dS/dlambdai=0}.
Thus \eqref{dS/dtn} is proved.

Equation \eqref{dS/ds} can be proved similarly. This completes the proof
of \thmref{thm:loewner->dcmkp}.

\section{Example}
\label{sec:example}

In this section we examine a simple example. Since we deal with the case
$N=1$, we omit the index of $\kappa_i$, $\lambda_i$, $F_i$ and so on.

Let $\alpha$ be a complex parameter. Direct computation shows that the
following function is a solution of the L\"owner equation
\eqref{r-loewner:g} for $N=1$:
\begin{equation}
 \begin{split}
    g(\lambda,z) &=
    -\alpha \frac
    {\sqrt{(1-\lambda\alpha z^{-1})(1-\lambda^{-1}\alpha z^{-1})} 
     + (1+\alpha z^{-1})}
    {\sqrt{(1-\lambda\alpha z^{-1})(1-\lambda^{-1}\alpha z^{-1})} 
     - (1+\alpha z^{-1})}
\\
    &=
    z \left(
         \frac{4\lambda}{(\lambda+1)^2} 
      - 2\frac{(\lambda-1)^2}{(\lambda+1)^2}\alpha z^{-1}
      + O(z^{-2})
    \right).
 \end{split}
\label{g:example}
\end{equation}
The driving function $\kappa(\lambda)$ is a constant function:
$\kappa(\lambda) \equiv \alpha$ and $\phi(\lambda)$ is determined by
\begin{equation*}
    e^{-\phi} = \frac{4\lambda}{(\lambda+1)^2}.
\end{equation*}
If $|\alpha|=1$ and $\lambda$ is a real number greater than $1$,
$\lambda\geqq 1$, the above $g(\lambda,z)$ is a conformal mapping from
\begin{equation*}
    \{z \in \Complex \mid |z|>1\} \smallsetminus
    \{t\alpha \mid 1 < t \leqq \lambda\}
\end{equation*}
to the outside of the unit disk, $\{w \in\Complex \mid |w|>1\}$.

The inverse function of $g(\lambda,z)$ is
\begin{equation}
 \begin{split}
    f(\lambda,w) &=
    \alpha \frac
    {(1+\lambda)(\alpha+w) +
     \sqrt{4\lambda(w-\alpha)^2 + (\lambda-1)^2 (\alpha+w)^2}}
    {(1+\lambda)(\alpha+w) -
     \sqrt{4\lambda(w-\alpha)^2 + (\lambda-1)^2 (\alpha+w)^2}}
\\
    &=
    w \left(
         \frac{(\lambda+1)^2} {4\lambda}
      + \frac{(\lambda-1)^2}{2\lambda}\alpha w^{-1}
      + O(w^{-2})
    \right).
 \end{split}
\label{f:example}
\end{equation}
Even in the case of $F(\lambda)=0$, this example gives a non-trivial
solution of the dcmKP hierarchy. As is seen from \eqref{f:example}, the
first Faber polynomial is
\begin{equation}
    \Phi_1(\lambda) = 
    \frac{(\lambda+1)^2} {4\lambda} w
    + \frac{(\lambda-1)^2}{2\lambda}\alpha.
\label{Phi1:example}
\end{equation}
Hence $v^1(\lambda) = \alpha (\lambda+1)^2/4\lambda$. If $t_n=0$ for
$n\geqq 2$, the hodograph relation \eqref{hodograph} becomes
\begin{equation}
    s + \alpha \frac{(\lambda+1)^2}{4\lambda} t_1 = 0.
\label{hodograph:example}
\end{equation}
Therefore
\begin{equation*}
   \frac{4\lambda}{(\lambda+1)^2} = - \frac{\alpha t_1}{s}, \qquad
   \frac{(\lambda-1)^2}{(\lambda+1)^2} = \frac{s + \alpha t_1}{s}.
\end{equation*}
Substituting them in \eqref{f:example}, we have
\begin{multline}
    \calL(s,t)|_{t_n=0 (n\geqq 2)}
\\
    =
    \alpha \frac
    {(\alpha+w)\sqrt{s} +
     \sqrt{-\alpha t_1 (w-\alpha)^2 + (s + \alpha t_1)(\alpha+w)^2}}
    {(\alpha+w)\sqrt{s} -
     \sqrt{-\alpha t_1 (w-\alpha)^2 + (s + \alpha t_1)(\alpha+w)^2}}.
\label{L:example}
\end{multline}

\section{Reduction to L\"owner-Kufarev equation}
\label{sec:loe-kuf}

In the theory of univalent functions, the L\"owner equation
\eqref{r-loewner:orig} is generalized in various ways. (See, e.g.,
\cite{dur:83} \S3.4.) Here we show that the {\em L\"owner-Kufarev}
equation (with multiple independent variables) of the type
\begin{equation}
    \frac{\der g}{\der \lambda_i}(\bflambda,z)
    = g(\bflambda,z) \frac{\der \phi(\bflambda)}{\der \lambda_i}
    \sum_{j=1}^N \mu_{ij}(\bflambda)
    \frac{\kappa_j(\bflambda) + g(\bflambda,z)}
         {\kappa_j(\bflambda) - g(\bflambda,z)}
\label{loe-kuf:g}
\end{equation}
turns into the L\"owner equation \eqref{r-loewner:g} by means of
suitable coordinate transformation. This gives another reduction of the
dcmKP hierarchy when it is combined with the result in
\secref{sec:loe->dcmkp}. 

The argument in this section is rather formal and we assume genericity
in many points.

We keep the assumption on the form of $g(\bflambda,z)$ \eqref{g(lam,z)}
as in \secref{sec:loewner}. Equation \eqref{loe-kuf:g} is rewritten
as the equation for the inverse function $f(\bflambda,w)$
\eqref{f(lambda,w)} as follows:
\begin{equation}
    \frac{\der f}{\der \lambda_i}(\bflambda;w)
    = 
    \left( 
    \sum_{j=1}^N
    \mu_{ij}(\bflambda) w
    \frac{w + \kappa_j(\bflambda)}{w - \kappa_j(\bflambda)}
    \right)
    \frac{\der \phi(\bflambda)}{\der \lambda_i}
    \frac{\der f}{\der w}(\bflambda;w).
\label{loe-kuf:f}
\end{equation}
By comparing the coefficient of $w^1$ (i.e., the asymptotic behavior at
$w=\infty$), it is shown that we need to assume 
\begin{equation}
    \sum_{j=1}^N \mu_{ij}(\bflambda)=1.
\label{sum-mu}
\end{equation}

Assume that $f$ is an entire function of $w$. Since the left hand side
of \eqref{loe-kuf:f} thereby does not have singularities at $w=\kappa_j$
($j=1,\dots,N$), we have
\begin{equation}
    \frac{\der f}{\der w}(\bflambda;\kappa_j(\bflambda)) = 0
\label{df/dw(kj)=0}
\end{equation}
for all $j$. Let us define new variables $\zeta_j$ as
\begin{equation}
    \zeta_j = \zeta_j(\bflambda) := f(\bflambda; \kappa_j(\bflambda)).
\label{def:zetaj}
\end{equation}
Assuming that the transformation
\begin{equation*}
    \bflambda=(\lambda_1,\dots,\lambda_N)
    \mapsto \bfzeta=(\zeta_1,\dots,\zeta_N)
\end{equation*}
is invertible, we show that \eqref{loe-kuf:f} becomes the L\"owner
equation \eqref{r-loewner:f} with respect to $\bfzeta$.

By \eqref{df/dw(kj)=0} and \eqref{loe-kuf:f}, we have
\begin{equation*}
 \begin{split}
    \frac{\der \zeta_j}{\der \lambda_i} &=
    \left.\frac{\der f}{\der \lambda_i}(\bflambda;w)
    \right|_{w=\kappa_j(\bflambda)}
\\
    &= \lim_{w\to \kappa_j(\bflambda)}
    \left( 
    \sum_{k=1}^N
    \mu_{ik}(\bflambda) w
    \frac{w + \kappa_k(\bflambda)}{w - \kappa_k(\bflambda)}
    \frac{\der \phi(\bflambda)}{\der \lambda_i}
    \frac{\der f}{\der w}(\bflambda;w)
    \right)
\\
    &= \lim_{w\to \kappa_j(\bflambda)}
    \left(
    \mu_{ij}(\bflambda) w
    \frac{w + \kappa_j(\bflambda)}{w - \kappa_j(\bflambda)}
    \frac{\der \phi(\bflambda)}{\der \lambda_i}
    \frac{\der f}{\der w}(\bflambda;w)
    \right)
\\
    &= 2 \mu_{ij}(\bflambda) \kappa_j(\bflambda)^2
    \frac{\der \phi(\bflambda)}{\der \lambda_i}
    \frac{\der^2 f}{\der w^2}(\bflambda;\kappa_j(\bflambda)).
 \end{split}
\end{equation*}
We can rewrite this equation as
\begin{equation}
    \mu_{ij}(\bflambda) 
    \frac{\der \phi}{\der \lambda_i}(\bflambda)
    = \frac{\der \zeta_j}{\der \lambda_i} K_j^{-1},
\label{mu-dphi/dlambda}
\end{equation}
if
\begin{equation}
    K_j:= 2 \kappa_j(\bflambda)^2 
    \frac{\der^2 f}{\der w^2}(\bflambda;\kappa_j(\bflambda))
\label{def:Kj}
\end{equation}
does not vanish. The L\"owner-Kufarev equation \eqref{loe-kuf:f} now
turns into the differential equation
\begin{equation}
 \begin{split}
    \frac{\der f}{\der \zeta_i}
    &=
    \sum_{j=1}^N \frac{\der \lambda_j}{\der \zeta_i}
    \frac{\der f}{\der \lambda_j}
    =
    \sum_{j=1}^N  \sum_{k=1}^N
    \frac{\der \lambda_j}{\der \zeta_i}
    \mu_{jk} w \frac{w + \kappa_k}{w - \kappa_k}
    \frac{\der \phi}{\der \lambda_j}\,
    \frac{\der f}{\der w}
\\
    &=
    \sum_{k=1}^N
    \left( 
     \sum_{j=1}^N \frac{\der \lambda_j}{\der \zeta_i}
     \frac{\der \zeta_k}{\der \lambda_j} K_k^{-1}
    \right)
    w \frac{w + \kappa_k}{w - \kappa_k}\,
    \frac{\der f}{\der w}
\\
    &= K_i^{-1}
    w \frac{w + \kappa_i}{w - \kappa_i}
    \frac{\der f}{\der w}
 \end{split}
\label{loewner:f:temp}
\end{equation}
with respect to $\zeta_i's$. This is almost \eqref{r-loewner:f} except
for the overall coefficient. Actually, if we compare the asymptotic
behavior at $w=\infty$, as we did when we derived \eqref{sum-mu}, we see
that
\begin{equation*}
    K_i^{-1} = \frac{\der \phi}{\der \zeta_i}.
\end{equation*}
Thus \eqref{loe-kuf:f} is rewritten as
\begin{equation}
    \frac{\der f}{\der \zeta_i}(\bfzeta;w)
    = w
    \frac{w + \kappa_i(\bfzeta)}{w - \kappa_i(\bfzeta)}
    \frac{\der \phi(\bfzeta)}{\der \zeta_i}
    \frac{\der f}{\der w}(\bfzeta;w),
\label{loewner:f:zeta}
\end{equation}
which is nothing but the L\"owner equation. 

Combining the above result with \thmref{thm:loewner->dcmkp}, we have the
following theorem.

\begin{theorem}
\label{thm:loe-kuf->dcmkp}
 Let $f(\bflambda,w)$ be a solution of the L\"owner-Kufarev equation
 \eqref{loe-kuf:f} of the form \eqref{f(lambda,w)} and $\bflambda(s,t)$
 be a solution of \eqref{hodograph}. Then the function
 $\calL=\calL(s,t;w)$ defined by
\begin{equation*}
    \calL(s,t;w) := f(\bflambda(s,t), w)
\end{equation*}
 is a solution of the dcmKP hierarchy \eqref{dcmkp}.
\end{theorem}

It is $\zeta_i(s,t)$ rather than $\lambda_i(s,t)$ that are the Riemann
invariants of this case. The functions $\lambda_i(s,t)$ satisfy the
equation
\begin{equation*}
    \frac{\der\lambda_i}{\der t_n}
    =
    \sum_{j,k=1}^N \frac{\der\lambda_i}{\der\zeta_j}
    v^n_j(\bflambda(s,t)) \frac{\der \zeta_j}{\der \lambda_k}
    \frac{\der\lambda_k}{\der s},
\end{equation*}
namely,
\begin{equation}
    \frac{\der\bflambda}{\der t_n}
    = Z^{-1} V_n Z
    \frac{\der\bflambda}{\der s}, \qquad
    \bflambda = {}^t (\lambda_1,\dots,\lambda_N),
\label{hydrodynamic;lambda}
\end{equation}
instead of \eqref{lambda(s,t)}. Here $Z$ and $V_n$ are $N\times N$
matrices defined by
\begin{equation*}
    Z = \left(\frac{\der\zeta_i}{\der\lambda_j}\right)_{i,j=1,\dots,N},
    \qquad
    V_n = {\operatorname{diag}}(v^n_1,\dots,v^n_N).
\end{equation*}

Ma\~nas, Mart\'{\i}nez Alonso and Medina \cite{m-a-m:02} considered the
chordal version (cf.\ \eqref{c-loewner:orig}) of the above
generalization. Our argument can be applicable to their case.

\section{Concluding remarks}
\label{sec:conclusion}

In this article we show that the dcmKP hierarchy can be reduced to the
L\"owner(-Kufarev) equations. Let us mention several further problems.

\begin{itemize}
 \item The dcmKP hierarchy is a ``half'' of the dispersionless Toda
       lattice hierarchy. Can we construct the missing half by extending
       the radial L\"owner equation? The dispersionless Toda equation
       has already appeared in \cite{g-m-a:03} but in a different
       context. 

 \item The stochastic version of the L\"owner equation and its
       application to the conformal field theories are intensively
       studied. See \cite{l-s-w:01}, \cite{car:03-05} and references
       therein. It might be interesting if the ``stochastic version'' of
       dispersionless type integrable systems (if any!) can be applied
       to the conformal field theories.
\end{itemize}

\end{document}